# Synopsis of the

# Workshop on Humor and Cognition

**Participants**

**Center for Research on Concepts and Cognition Members:**

Douglas Hofstadter (Professor of Cognitive Science and Computer Science, Indiana University) and graduate students David Chalmers, Robert French, Liane Gabora, Melanie Mitchell, and David Moser. Also Peter Suber (Professor of Philosophy, Earlham College) and Kevin Kinnell (programmer). Other Indiana University Faculty Members: Igor Gavanski (Psychology), Steven Sherman (Psychology), and Allen Winold (Music Theory).

**Visitors:**

Victor Raskin (Professor of English, Purdue University) and graduate students Salvatore Attardo, Donalee Hughes, Maureen Morrissey, and Yan Zhao. Also Charles Brenner (programmer), Scott Buresh (lawyer), Gray Clossman (programmer), and Richard Salter (Professor of Mathematics and Computer Science, Oberlin College).

## ABSTRACT

In this paper we summarize the proceedings of the Workshop on Humor and Cognition held at Indiana University's Center for Research on Concepts and Cognition on February 18 and 19, 1989. The principal type of humor considered, slippage humor, is first defined and contrasted with aggression-based humor. Next, a particularly clear variety of slippage humor, based on Douglas Hofstadter's notion of a frame blend, is presented. Given that a frame is a small coherent cluster of concepts pertaining to a single topic (similar to Victor Raskin's notion of a script), a frame blend is what results when elements are extracted from two distinct frames and spliced together to yield a new hybrid frame. Diverse ways of blending two given frames can produce varying amounts and types of humor, and some studies of this phenomenon are presented. A close connection between frame blends and analogies is pointed out. To make this connection more explicit, the Copycat domain — an idealized microworld in which- analogy-making can be studied and modeled on computer — is presented, and it is shown how jokes can be mapped into that domain, giving rise to a kind of abstract "microworld humor". The reduction of these phenomena to the Copycat domain helps to bring out the tight relationships among good jokes, defective analogies, and frame blends quite clearly. As a result, these relationships appear clearer in the real world as well. The notion that many jokes can share the same abstract structure is suggested, and the name ur-joke is suggested for the most abstract level of a joke. Several specific ur-jokes are presented, each one with a set of fully fleshed-out jokes based on it. We recount the group's collective efforts at translating two jokes from one subject matter to another, in an attempt to determine whether a joke's funniness is due more to its underlying- ur-joke or to its subject matter. This important question is, however, left open. There follows some discussion of Victor Raskin's overlapping-script theory of humor, which has many points of contact with Hofstadter's frame-blend theory, and then a summary of Salvatore Attardo's theory of a multiple-level analysis of jokes (closely related to Hofstadter "ur-joke hypothesis") is presented. Finally, a speculative theory by Gray Clossman about the adaptive value of humor is briefly addressed.

## Introduction: Raison d'être of the Workshop

At the Center for Research on Concepts and Cognition, we are using computer models to work toward an understanding of the mechanisms underlying human thought. Fundamental to cognition is the fact that human concepts exhibit fluidity, that is, each concept overlaps with, and has a propensity to "slip" into, numerous other concepts, depending on mental pressures evoked by the current context. For example, humans routinely come up with counterfactuals — hypothetical variants of an actual event that involve slippages from the true way it happened to ways that it might otherwise have happened. Careful study and modeling of such conceptual slippages (not to be confused with slips, or errors) helps to reveal how concepts are organized.

To study counterfactuals is one way of investigating the fluid nature of human concepts; to study analogy is another. We view analogy-making as the process of recognizing that two quite different-seeming situations are actually the same, as long as certain slippages are made. Another way of putting it is to say that the situations share a single "conceptual skeleton", or essence. In most analogies of any complexity, slippages come at various levels of abstraction, exhibit varying degrees of "tension" or "stretch", and run in families. Any model of analogy that faithfully reflects all these subtleties must certainly incorporate a deep understanding of conceptual fluidity and slippability. We believe, therefore, that research into analogy will provide much insight into how people categorize objects and situations, construct counterfactuals, are reminded by one situation of another, and come up with creative insights (Hofstadter 1985, chapters 12 and 24).

One of the ongoing projects at CRCC is the development of Copycat, a computer program that solves idealized analogy problems involving strings of letters (Hofstadter, 1984; Hofstadter and Mitchell, 1988; Mitchell, 1988). Each Copycat analogy problem has numerous answers of varying degrees of plausibility, and in studying the less plausible answers to certain problems, we discovered, somewhat to our surprise, that some of them provoked laughter in much the same way as certain jokes do. In fact, once we had noticed this connection, we were able to make rather tight correspondences between some Copycat analogies and specific jokes, which made us aware of how closely analogy and humor are related.

The Workshop on Humor and Cognition was therefore motivated to a large extent by the observation that jokes have much in common with analogies gone awry, and by the belief that through exploration of the similarities and differences between humor and analogy, we would sharpen our understanding of both processes, and of the fluid nature of human thought in general.

## Slippage Humor and Frame Blends

The workshop began with a talk by Douglas Hofstadter, in which he defined and discussed what he cplls "slippage humor". As a prototype of this notion, he offered the following casual remark made by David Moser while wandering in Harvard Square near the music store Briggs & Briggs:

> If Harvard Square were Harvard Cube, Briggs & Briggs would be Briggs & Briggs & Briggs!

Note that this is both a joke and a counterfactual. The first slippage here is from two dimensions to three, and the second slippage, conceptually parallel to the first (at least on a superficial level), is from two copies of the name "Briggs" to three.

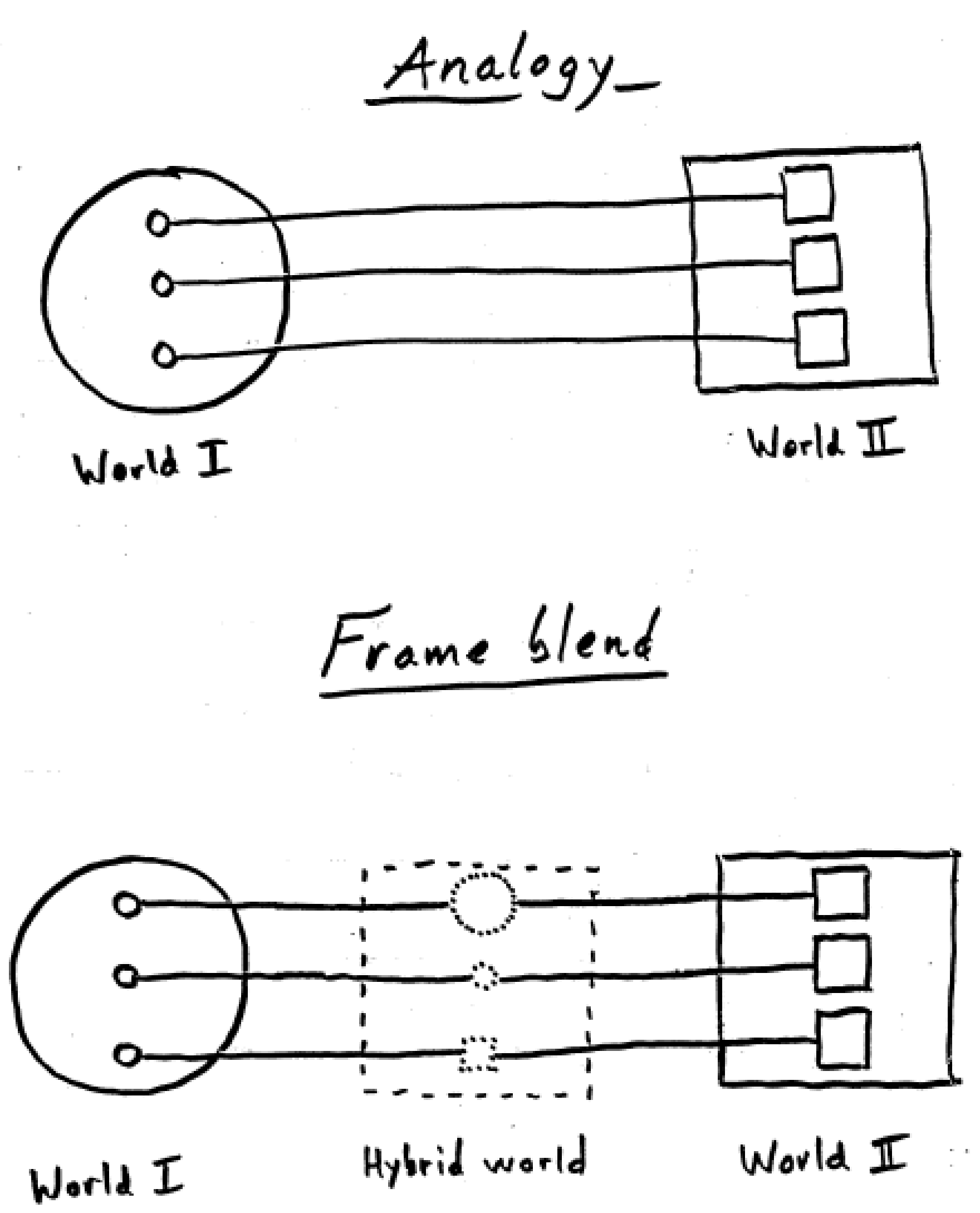

Figure 1. A schematic comparison of an analogy (above) and a frame blend (below).

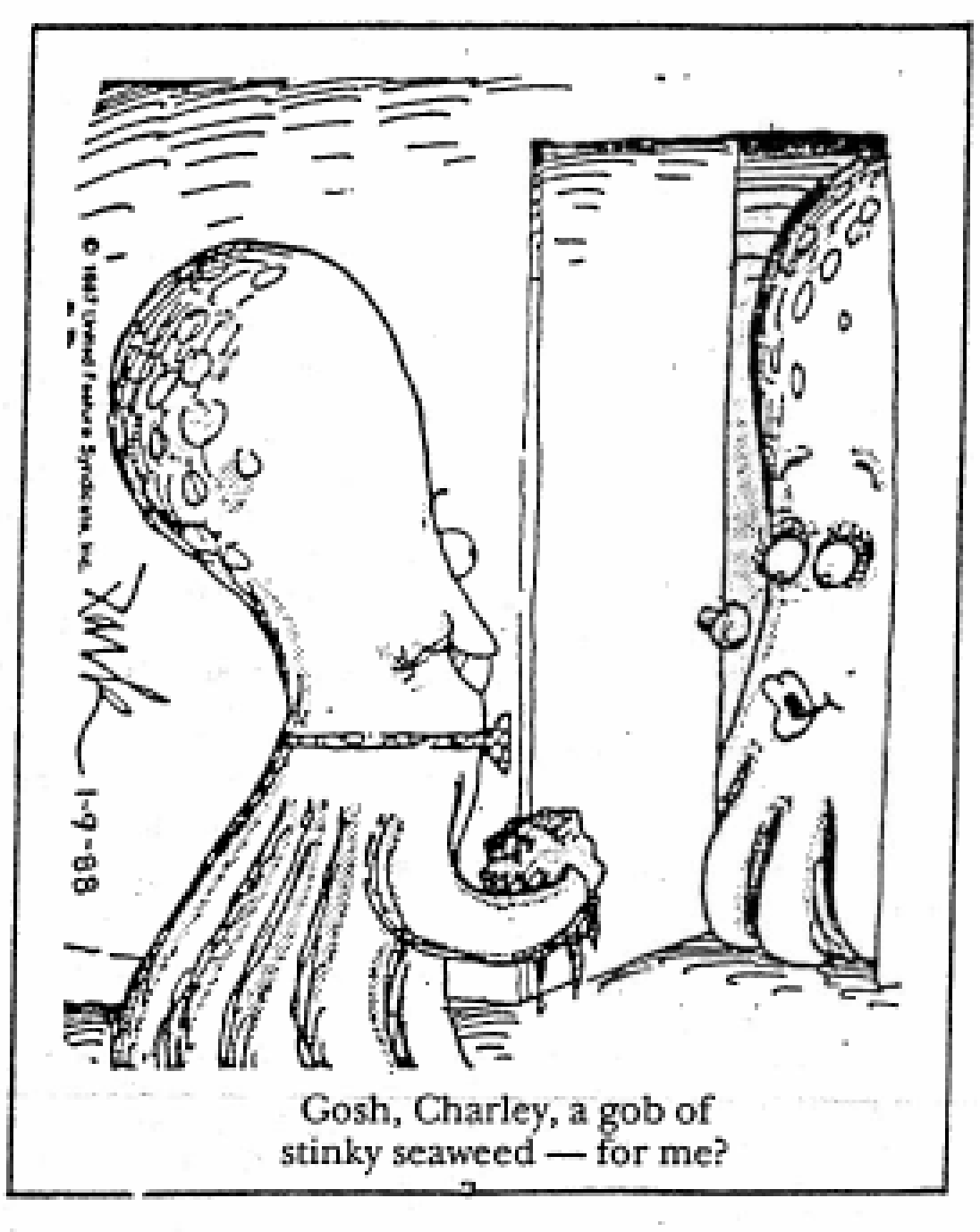

Figure 2. An Off the Leash cartoon by W. B. Park that blends a human-courtship frame with an octopus frame.

Clearly, this joke operates at a purely cognitive or intellectual level, and it would be very hard to see it as being a "safe outlet for aggression", despite the claims of numerous authors on humor (Sigmund Freud, Henri Bergson, and Arthur Koestler, among others) that all humor is based on aggression. Hofstadter argued that, quite to the contrary, humor can be completely innocent, and, as this joke shows, can simply derive from a bizarre combination of slippages.

Having defined slippage humor as his focus, Hofstadter then turned to a particular form of it — namely, humor based on frame blends. A frame blend occurs when a person blurs two distinct situations or scenarios in their mind, creating a hybrid situation composed of aspects of each situation. Certainly not all frame blends are incoherent, let alone funny — yet many are. Frame blends are closely related to analogies, for the simple reason that people will not confuse two situations unless one reminds them of the other — and this happens only when the two situations are analogous at some level. The act of constructing an analogy establishes many counterpart relations between the frames. If a and b are counterpart objects in frames A and B, respectively, then most typically, if a is incorporated into the blend, b will not be, and vice versa. Often, one frame will predominate over the other, and just one or two elements of the lesser frame will slip into the hybrid frame. Figure 1 schematically portrays the relationship between analogies and frame blends.

Hofstadter observed that a very popular contemporary style of humor is based on the blending of human and animal frames, as in Gary Larson's Far Side cartoon series and W. B. Park's Off the Leash series. He displayed as an example an Off the Leash cartoon in which elements from a human-courtship scenario are blended with elements from an octopus scenario (Figure 2). Figure 3 attempts to spells out more• explicitly how the two frames are blended, and in addition indicates (via wiggly lines) the analogy upon which the blend is based.

One can consider the two frames in a frame blend to be related as are figure and ground in a piece of visual art. The dominant frame (playing the role of ground) defines a context against which imported elements of the lesser frame stand out (thus acting as figure). For example, in Figure 2, the role of ground is played by the human courtship frame. Octopuses have been imported into it and catch our attention (as does any figure against a ground), which is why, on first glance, one might think that the octopus frame is the dominant one. Each octopus constitutes a subframe in its own right, and as such is open to further blending. That is, each octopus acts as a second-order "ground" onto which a "figure", borrowed from the human frame,_ can be painted. Specifically, the introduction of lipstick and eye shadow onto the female octopus, and a bow tie onto the male octopus, are "second-order" frame blends. (If there were a picture of an octopus on the bow tie, that would constitute a third-order frame blend — and so on.)

A particularly important role is played by the words "stinky" and "gob", which not only have been borrowed from the human frame (octopuses, after all, do not talk!), but also reflect a distinctly human attitude toward seaweed. Supposing that octopuses could talk, one imagines that the female's words of thanks would employ terms exuding positive connotations — the analogues, or translations, of human phrases about flowers. Thus, one can more realistically imagine her saying, in Octopese, something like "Gosh, Charley, a twist of fragrant seaweed — for me?" What Park has done to this hypothetical utterance is a kind of illegitimate back-translation into English, taking into account the fact that whatever term expresses fragrance in Octopese applies to objects that humans consider malodorous. It is this twist that gives the cartoon much of its pizzazz.

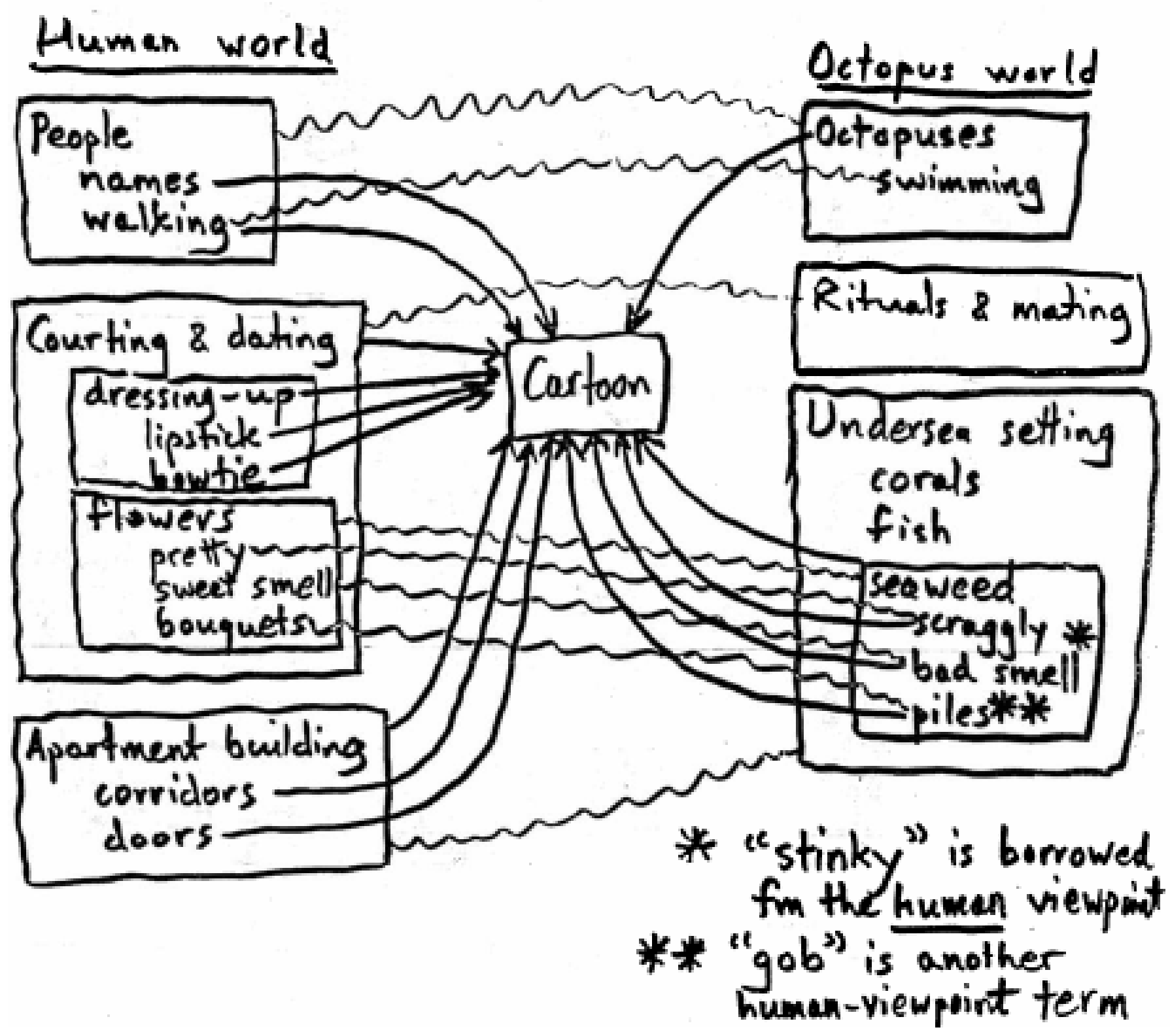

Figure 3. Spelling out how the two frames in the cartoon in Figure 2 are blended, and indicating (via wiggly lines) the analogy upon which the blend is based.

One can consider any frame blend as a theme upon which to make variations by adjusting the relative amounts of its component frames. Often, there are several distinct conceptual dimensions that can be varied independently, and these can be thought of as "knobs" that can be "twiddled". For example, in the cartoon in Figure 2, a few possible "knobs" might be: the species of each individual, the object proffered by the male, and the words of gratitude uttered by the female. Kevin Kinnell has made a set of variants of this cartoon by letting each of these knobs take on a few discrete settings, as follows:

Species of male: human vs. octopus
Species of female: human vs. octopus
Object proffered: flowers vs. seaweed
Connotations of words: positive vs. neutral vs. negative

With these knob settings, the number of possibilities is 2 x 2 x 2 x 3, or 24 in all. Four interesting and fairly complementary representatives of this family of variants are shown in Figure 4.

It is not always the case that frame blends combine many elements from each frame. In some frame-blend jokes, a thorough blend, as well as a high degree of humor, is achieved very economically by the importing of a single but telling element from an alien frame into an otherwise intact frame. A good example of this is the Off the Leash cartoon shown in Figure 5, in which the scenario of pig-feeding is violated by the single French word "Garcon!", eloquently conjuring up a vivid French-restaurant scenario.

Eric Haas, a former student of Hofstadter's, carried out a study of knob-twiddling on this frame blend, and came up with several interesting variants, shown in Figure 6. One not shown, but perhaps the funniest of all (including the original), features the original scene with the variant caption, "Garcon, there's an escargot in my slop!" — an obvious allusion to the famous line, "Waiter, there's a fly in my soup!"

When one sees such a family of variants of a particular cartoon, one cannot help but wonder if any of these images or ideas (or related ones) flashed through the cartoonist's mind at either a conscious or an unconscious level, and whether the cartoonist would find any of the alternates funnier than the "official" version.

## Copycat Analogies, Frame Blends, and Jokes

As was mentioned previously, we are developing a computer program, Copycat, that makes analogies between idealized situations consisting of letter strings. A sample problem from the Copycat domain is the following:

If the string **abc** is changed to **abd**, how can one change **ijk** "the same way"?

Because all humans share an evolutionary history and also have fairly similar sets of experiences, we all tend to perceive structure in similar manners, and thus give similar answers to problems of this sort. In this particular problem, most people view the initial event as "replacement of the rightmost letter by its successor". Straightforward application of this rule to the target string **ijk** yields **ijl**. It would be possible, nonetheless, to take the change much more literally — namely, as "replacement of the rightmost letter by **d**" — and thus to answer **ijd**. Few people see this as a better answer than **ijl** — in fact, few people even think of it at all. An even more literal-minded (and thus far-fetched) answer would be **abd**, which would be based on seeing the initial event simply as "replacement of the entire string, lock, stock, and barrel, by **abd**".

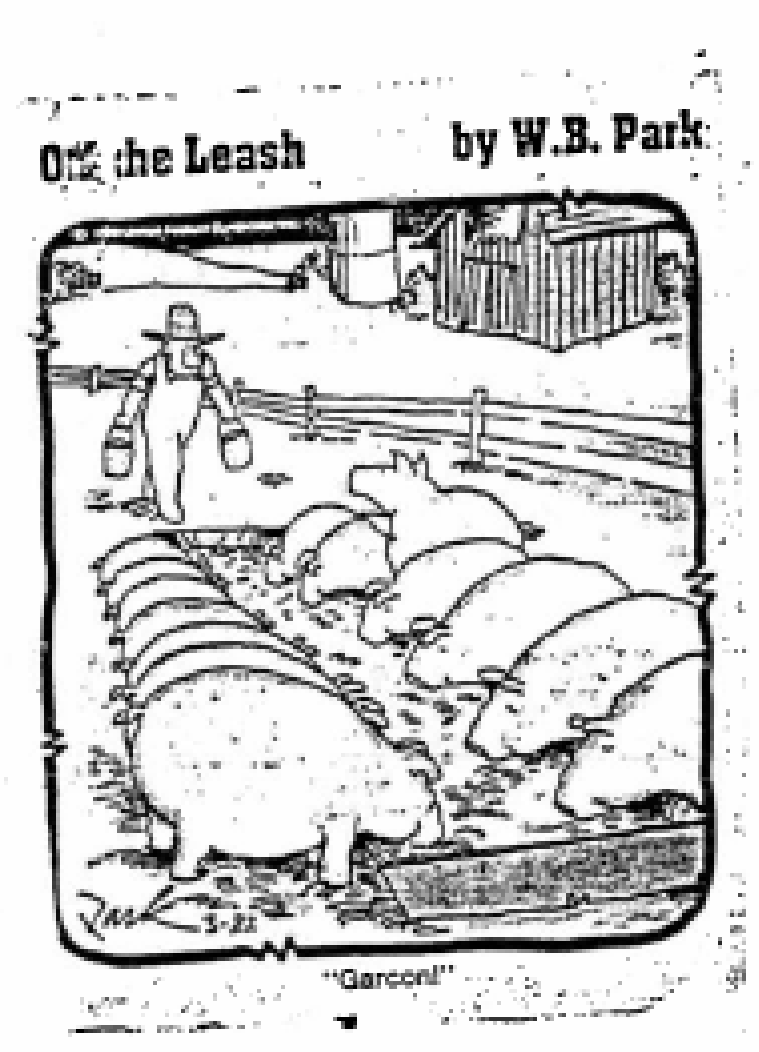

Figure 5. Another Off the Leash cartoon by W. B. Park, which imports a single element from a French-restaurant frame into a pig-feeding frame.

This is so brutally simplistic that virtually no one ever suggests it, even in jest. Natural selection has clearly favored the evolution of brains that automatically perform some degree of abstraction.

The answer **ijd** is a simple frame blend, in which the **abc/abd** frame contributes just one element — the **d** to the **ijk** frame. Seen this way, this answer to the problem bears a strong similarity to the following well-known joke:

> *American:* Look how free we are in America — nobody prevents us from parading in front of the White House and yelling, "Down with Reagan!"
> *Russian:* We in Russia are just as free as you — nobody prevents us from parading in front of the Kremlin and yelling, "Down with Reagan!"

Here, the Russian attempt "translate" the notion of free speech from an American to a Soviet frame, but instead of carrying it fully across (as would happen in a good analogy), blurs frames by importing Reagan literally into the Soviet frame. Thus a bad analogy, in the form of a frame blend, makes for a good joke. One can also model a joke on the even dumber answer **abd**, as follows:

> *American:* Look how free we are in America — nobody prevents us from parading in front of the White House and yelling, "Down with Reagan!"
> *Russian:* We in Russia are just as free as you — nobody prevents us from parading in front of the White House and yelling, "Down with Reagan!"

This is not only a pathetic analogy, it is also a rather feeble joke, thus defeating the optimistic but simplistic theory that any bad analogy will give rise to a good joke. One might well ask for a characterization of just which bad analogies will make good jokes, and why, but unfortunately those questions are far from answerable at this time. For-the sake of completeness, let us also show how the good answer iji translates into the US-SU situation:

> *American:* Look how free we are in America — nobody prevents us from parading in front of the White House and yelling, "Down with Reagan!"
> *Russian:* We in Russia are just as free as you — nobody prevents us from parading in front of the Kremlin and yelling, "Down with Gorbachev!"

As is evident, since the analogy has been carried out perfectly, there is no humor whatsoever in the Russian's remark.

Hofstadter's primary focus in mapping out the connection between bad analogies and jokes was a different and somewhat richer Copycat analogy problem, namely:

> If **abc** is changed to **abd**, how to change **xyz** "the same way"?

A short discussion of this problem is in order before we can discuss its relation to jokes. Most people start out by attempting to replace the rightmost letter by its successor, but since **z** has no successor, this poses a problem. Many people then invoke the concept of circularity and produce the answer **xya**. However, circularity was deliberately excluded from the Copycat microworld in order to force analogy-makers to restructure their original perception of the strings, and hopefully to discover new, insightful slippages. Hofstadter dubbed the impasse that analogy-makers find themselves in at this point "the snag".

Goaded by this snag, some people notice that **xyz** is "wedged" against the end of the alphabet, and that **abc** is symmetrically wedged against the beginning. This symmetric opposition suggests that the **a** and the **z** be seen as counterparts. Symmetry further suggests that the c and the x be seen as counterparts. Once one has created this "reversed" mapping between abc and xyz, one sees it is appropriate to slip the concept "successor" to its opposite as well — namely, the concept predecessor". Together, this set of conceptually parallel slippages yields the answer wyz — an elegant answer, as well as a clever way of getting around the snag.

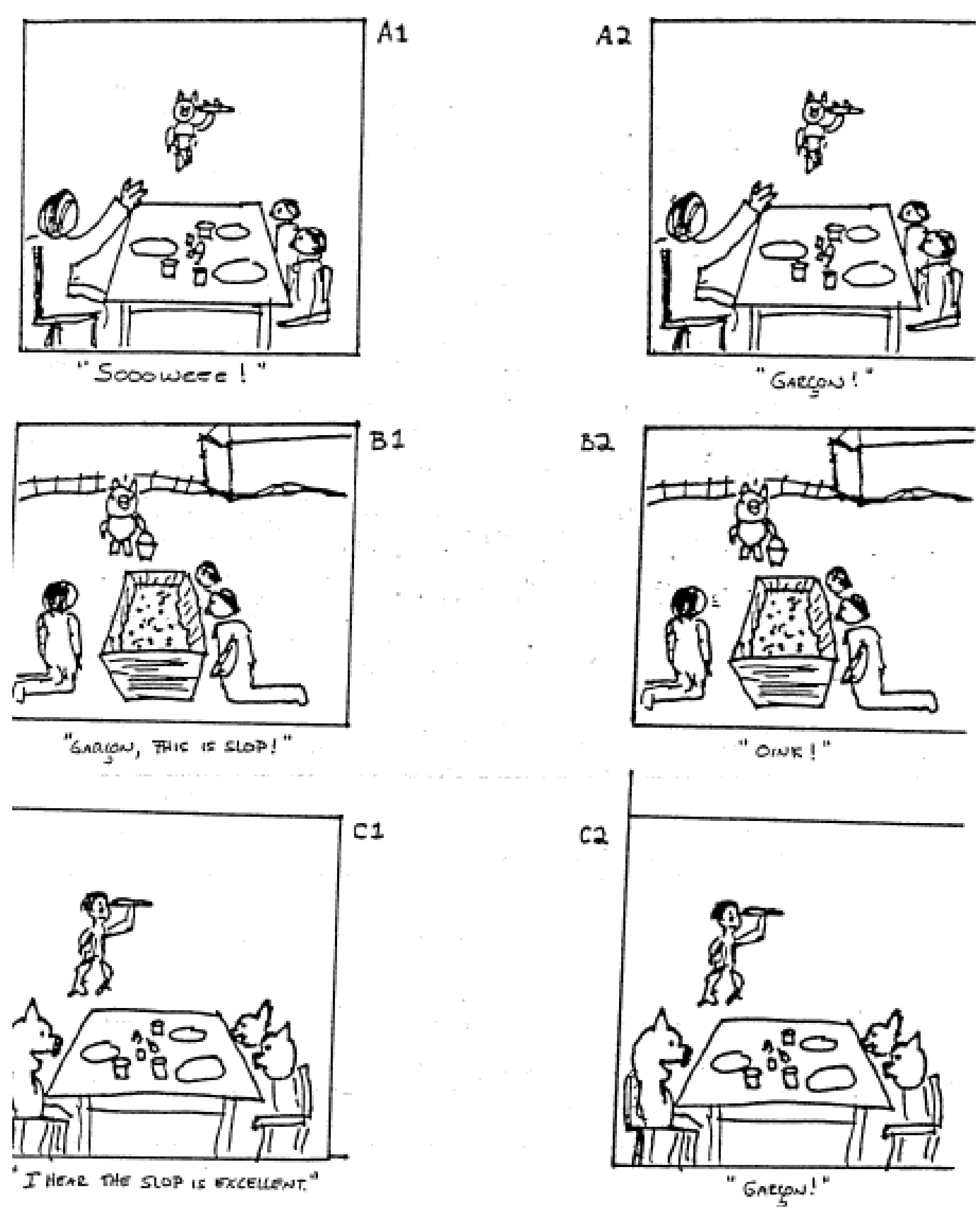

Figure 6. Variations by Eric Haas on the Cartoon in Figure 5.

Interestingly, however, relatively few people actually come up with **wyz**. Many people give answers based on the tacit assumption that z is the element in the target that should be modified. These include **xyd** (same rationale as for **ijd**, except that here, it seems more justified, given the snag), **xy** (sometimes pronounced "**xy**-blank"), and **xyy** (go to the end of the alphabet and bounce backwards). Although each of these answers can be justified, nonetheless most people agree, once they have seen the possibilities, that **wyz** is the most esthetically satisfying answer.

In our exploration of alternative answers to the **xyz** problem, we have discovered several solutions that strike many people as quite funny. The simplest (and probably most likely to be actually suggested by a human being) is the "dizzy" answer **dyz**. In this answer, the rather abstract symmetry between **abc** and **xyz** has been insightfully perceived, thus suggesting the **x** as the letter to modify. However, the hypothetical proposer of this answer falls prey to the very problem — having to rigidly export the letter **d** — that they went to great lengths to avoid! Why not simply say **xyd**, if you are going to export the **d**?

An even dizzier answer is **dba**, based on combining the super-rigid answer abd with the highly sophisticated observation of the symmetry of the **abc** and the **xyz**. Here, the symmetry is taken into account by writing **abd** backwards — as if that somehow makes it no longer dumb.

A final example of dizziness is the answer **abw**, which is a different way of combining tremendous insight with tremendous rigidity. Here, the solver has apparently seen all the way to the idea of replacing **x** by **w**, and yet instead of doing so, has replaced the **c** — the **x**'s counterpart inside **abc** — by **w**. Talk about confusion!

In fact, all these "scramble-brained" answers are fusions of great insight with pathetic stupidity, reminding us of a football player who makes a beautiful catch of a long pass and then fumbles the ball on the one-yard line. This abstract skeleton is shared by many jokes, but we will let the following one (due to Johnny Carson) serve as a representative:

> Nancy Reagan insisted on the free distribution of the Government butter surplus to the truly needy, saying, "Even these poor people must have *something* to dip their lobster Inns into."

Our poor, dizzy First Lady initially gives the impression of knowing that to be poor is to be needy, but her closing words reveal that she has no understanding of poverty at all.

## The Notion of an Ur-Joke

The idea, just mentioned, of an abstract skeleton shared by many different jokes, was the next major topic that Hofstadter addressed. He pointed out that in humor there are certain recurrent themes, or joke-skeletons, upon which many jokes are built, much as in mathematics, certain abstract structures crop up over and over in the proofs of theorems. He termed such a skeleton an urjoke. (The prefix "ur" comes from German, and means "original" or "primordial".) The concept of an ur-joke is similar to the music-theoretical notion of an "Urlinie", invented early this century by Heinrich Schenker. The Urlinie of a given piece of music is supposed to be its deepest melodic core, arrived at by repeated stages of stripping away ornamentation, until only a few notes are left. Just as many pieces of music can share the same Urlinie, differing merely in how they "clothe" it, many very different jokes can share the same ur-joke, differing merely in how they "clothe" it — that is, what setting they place it in, how the joke is phrased, and so on. Complicating matters, however, is the fact that many jokes are built on a combination of ur-jokes, much as several different mathematical ideas are often combined in the proof of a theorem.

To exemplify the ur-joke notion, Hofstadter chose the theme of role reversal, in which there is a switching of normal or default roles or concepts (schematized in Figure 7). Here is a typical joke based on the "role reversal" ur-joke:

Question: What did Mickey Mouse get for Christmas?
Answer: A Dan Quayle watch.

Two other good examples sharing the same ur-joke are the cartoons by Gary Larson and Ed Fisher, shown in Figure 8.

The theme of role-reversal can be realized very crisply in the Copycat domain. Consider the problem:

If **abc** is changed to **abs**, how to change **pqr** "the same way"?

Before one sees the target frame, **pqr**, the letter s seems arbitrary. Indeed, if the target were tilt, the answer **ijs** would be as good an answer as one could hope for. However, once one has seen the target, the letter s can be given a very simple justification: it is the successor of the target s rightmost letter! Thus in the initial event, an element has been exported from the **pqr** frame into the **abc** frame. In order to mirror this effect in the second frame, a reverse borrowing should take place — namely, one should export from the **abc** frame the successor of its rightmost letter. This then yields the rather surprising and intriguing answer **pqd**. In fact, there is a degree of humor to this answer, despite the fact that it is a good analogy rather than a bad one.

Is it possible to copy this Copycat analogy, and to find a joke that is very close to it in flavor? A candidate for that role is this sign, occasionally found posted at private swimming pools:

We don't swim in your toilet — please don't pee in our pool!

It is worthwhile pointing out that, since all these role-reversal jokes share, at their core, a single conceptual skeleton, they are all united by analogy. Another quite common ur-joke is the "almost"-situation (Hofstadter 1979, pp. 634-643). Here are two humorous anecdotes that share the same skeleton:

> Liane Gabora recounted how, on her way through a cemetery, she passed a gravestone with the name "Norma Joan Baker" engraved on it. "Wow," exclaimed Bob French, "Marilyn Monroe is almost buried there!" [One has to know that Marilyn Monroe's original name was "Norma Jean Baker".] Bil Lewis, at the time a student of Douglas Hofstadter, once remarked, 'My uncle was almost President of the US!" "Really?" said Hofstadter incredulously. "Sure," replied Bil, "he was skipper of the PT 108!" [Again, one has to know that John F. Kennedy was skipper of the PT 109.]

Self-undermining comments, such as "I'll stop procrastinating tomorrow" and "I've told you a million times not to exaggerate", form another class of jokes all sharing a single ur-joke. Some jokes in this category, such as "Thank God I'm an atheist!" and "I'd give my right arm to be ambidextrous", contain metaphors that backfire when interpreted literally.

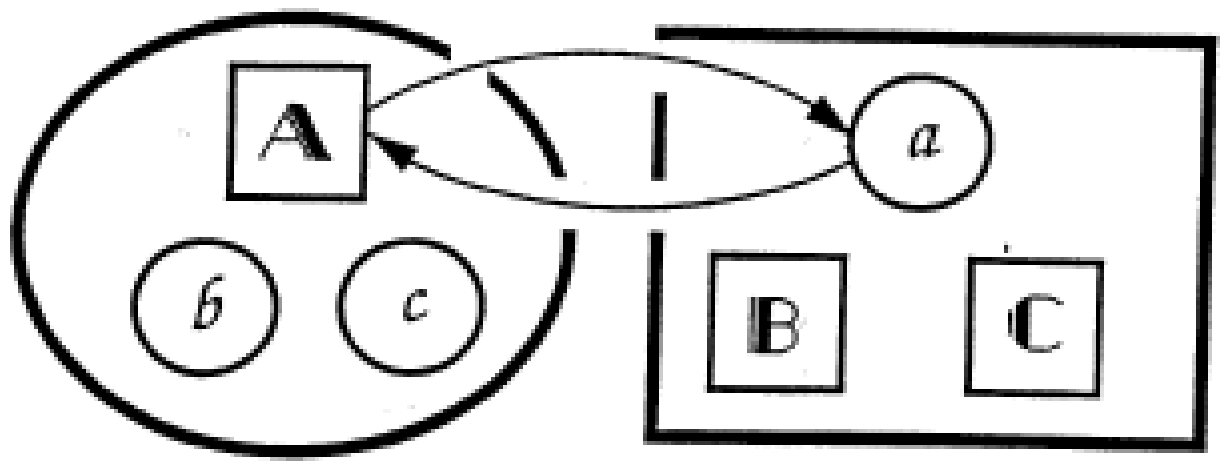

Figure 7. A schematic representation of the role-reversal ur-joke.

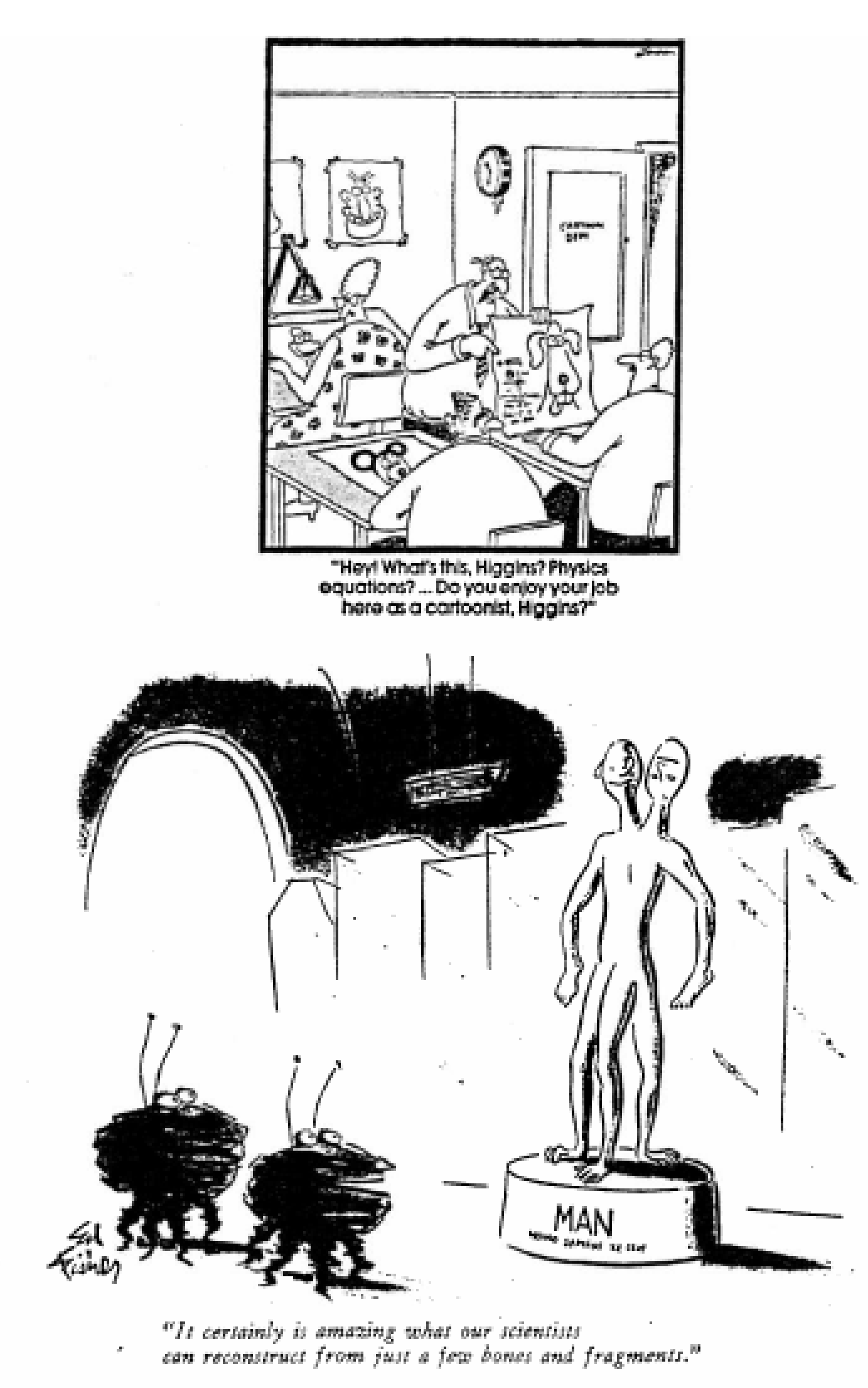

Figure 8. Role-reversal cartoons by Gary Larson and Ed Fisher.

Though all members of an ur-joke family share a conceptual skeleton, they certainly can differ in degree of funniness. Hofstadter srculated that this is perhaps due to differing degrees of difficult)/ in "unmasking the ur-joke — that is, in recognizing which familiar ur-joke is hidden in the complex wrapping. For example, some people feel that "I've told you a million times not to exaggerate" is not nearly as funny as "I'd give my right arm to be ambidextrous"; perhaps this discrepancy is due to the fact that it takes a moment longer to detect the self-undermining quality in the latter example. One might argue that, much as the shelling of walnuts contributes to the pleasure of eating them, the process of unmasking of a joke enhances our enjoyment of it. In summary, then, the theory that was suggested is that if two jokes sharing the same skeleton are not equally funny, the reason must be that the skeleton is buried to different amounts in the different jokes.

## How Deep Lies the Essence of a Joke?

Some members of the group took issue with this theory, arguing that certain topics of discourse — sex, of course, but also death, religion, politics, ethnic groups, and so on — have inherent tension associated with them, and that much of the humor of a particular joke is due not so much to its ur-joke, but to its subject matter. This thesis, diametrically opposite to the one Hofstadter proposed, gave rise to an interesting challenge. It was proposed that the group attempt to "translate" two sexual jokes into equally funny nonsexual jokes, preserving the ur-joke, of course. The first joke tackled was this one:

> A man in his fifties goes to the doctor and says, "Doc, I've got a problem. You see, when I was younger I always used to get erections that I couldn't bend with my hand. Now though, I can bend every erection I get. What I want to know is, am I getting stronger or weaker?"

Liane Gabora proposed the following nonsexual translation:

> A woman goes to the psychiatrist and says, "Doctor, I've got a_ problem. You see, when I was younger I loved making puzzles for myself and then trying to solve them. It used to be that the puzzles I invented were so difficult that I couldn't solve any of them. These days, however, I solve every puzzle I make up. The question is, am I getting smarter or stupider?"

The group felt that this version was a moderately successful translation of the joke, but that it wasn't as funny as the original. Douglas Hofstadter observed that both Gabora's translation and the original joke reminded him of the old paradox about God — namely, "Can God make a stone so heavy that God can't lift it?" Scott Buresh immediately took that hint and adroitly converted it into the following translation:

> God goes to the doctor and says, "Doc, I've got a problem. You see, I used to be able to make stones that were so heavy that I couldn't lift them. But now I can't make a stone that I can't lift. The question is, am I getting more or less divine?"

Someone observed that "omnipotent" might work better than "divine" (and indeed, it slyly recalls the impotence of the man in the sexual version), and the group agreed. Thus the final translation became this:

> God goes to the doctor and says, "Doc, I've got a problem. You see, I used to be able to make stones that were so heavy that I couldn't lift them.

> But now I can't make a stone that I can't lift. The question is, am I getting more or less omnipotent?"

Curiously, this joke elicited considerably more laughter than did the original joke. One possible reason for this is the fact that it was a translation. A second possible reason is that, for many people, the opening five words evoke a funny scenario all by themselves. Still, the fact remains that we were able to translate this joke with success out of the realm of sex.

Having seen "the same joke" in three different domains, we can attempt to verbalize explicitly the ur-joke that lies beneath. Here is a first pass:

> An individual interprets two tests as measuring the same desirable quality (strength, intelligence, etc.). Not only are different answers obtained on the two tests, but in fact, a high score on one test necessitates a low score on the other; they are merely flip sides of each other! Then at a later time, the outcomes of the same two tests are reversed, but of course this is of no more diagnostic value than the first time. So the individual remains baffled.

Summarized even more concisely, the ur-joke is this:

> Someone already confused by a double-edged message becomes even more stymied when confronted by its opposite, an equally double-edged message.

Perhaps this is too concise and omits something crucial; on the other hand, perhaps it can be further refined or rendered clearer in some sort of diagram. In any case, this set of abstract relationships — give or take a little — forms a conceptual skeleton that can be realized in different domains. There may even be a number of jokes already in circulation that share this skeleton.

The second joke that the group attempted to "de-sex" was the following one:

> As the Grande Finale to his act, the circus lion-tamer sits his fiercest lion in the center of the ring, with its snarling mouth wide open. But instead of placing his head inside the beast's gaping jaws, the lion-tamer unzips his trousers and uses his tool. The crowd gasps in amazement, and after a full minute has passed, the lion-tamer withdraws and puts it away. He then offers $500 to anyone in the audience who will do the same. Not a murmur comes from the audience, so the lion-tamer increases his offer to $600, then $700, and finally to $1,000. At this point, a small chap in the audience comes forward and says he will do it for $1,000. The crowd is stunned, and the lion-tamer warns him that it is extremely dangerous. "Are

> you absolutely sure that you want to go through with this?" he asks. "Well," says the small chap, "I'll do my best, but I'm not sure if I can open my mouth as wide as the lion's."

Gray Clossman came up with the following nonsexual way of telling this joke. An archer shoots an apple off an assistant's head and then asks for volunteers. The volunteer's response implies that she is willing to aim at an apple perched on the archer's head. The group felt that the essence of the joke is not fully preserved in this form, however, since the person who asks for volunteers is not the person who takes the risk. More importantly, the translated version doesn't seem as funny as the original, perhaps because the image of a man putting his erect penis in a lion's gaping mouth is very weird and ludicrous, and has no counterpart in the archery version. The group did not come up with any better translation of this joke, but on the other hand, it did not work very long at this second challenge.

From these two very cursory translation exercises, the group was unable to draw a clear conclusion about the issue in dispute — namely, the extent to which a joke owes its funniness to its ur-joke alone. On this subject, it is an interesting fact, although possibly a misleading one, that when an ur-joke is presented "bare" — either verbalized or diagrammed — it seems to possess absolutely no humor on its own. On the other hand, there is a noticeable difference in degree of humor between a "bare" ur-joke and one wearing very summery Copycat clothes. This reveals that to provoke laughter, an ur-joke definitely needs some clothing, but the clothing can be surprisingly scanty. Apparently the act of "disrobing" the ur-joke from clothing, no matter how skimpy, is sufficiently interesting to yield some humor.

This as-yet unresolved issue — "Does humor derive from the ur-joke or from the domain?" — is, in a sense, humor's analogue to the nature-vs.-nurture debate in child-rearing, where the role of genes is played by the ur-joke, and the role of the environment is played by the domain of discourse. We hope, however, that this issue will not remain as intractable as the nature-vs.-nurture controversy!

## Adaptive Functions of Humor

Toward the end of the workshop, there was some discussion of how humor might have evolved and what adaptive functions it might serve. Gray Clossman sketched out a theory according to which jokes coax us to play with and map onto one another the frames they implicate, and that this activity leads us to a better understanding of the nature and boundaries of these frames. This in turn leads to an increased understanding of, and thus ability to function in, the world around us. Furthermore, as was suggested by Peter Suber, since the skills involved in creating and understanding humor are also involved in other acts of intelligence and creativity, humor functions as a mental exercise; it prepares our minds to deal flexibly with cognitive tasks.

## A Few Qvick Personal Notes on Humor

by Doug Hofstadter
February, 1989

(1) Humor — at least what I mean by the term — is largely a cognitive or intellectual phenomenon, rather than an emotional one (e.g., a "safe" outlet for aggression, as such authors as Freud, Bergson, Koestler, et al. have maintained). Humor is usually provoked by some combination of surprise, delight, and absurdity.

(2) There are certain recurrent themes in humor, much as in mathematics. These could be called ur-jokes (like Heinrich Schenker's notion of "Urlinien" in music). Thousands of distinct jokes are built on each given ur-joke — they differ in how the ur-joke is "clothed" or "wrapped", or possibly distorted.

(3) When one knows for sure that a joke is coming (as when one looks at a cartoon or listens to a joke-teller telling a story), much of the pleasure induced by "getting" the joke is in the act of recognition of the theme — that is, recognizing which familiar ur-joke it is. This is analogous to the pleasure experienced by a child unwrapping a birthday present — the unveiling is a very major component of the event.

(4) Often, one or more ur-jokes are combined in some unexpected way, much as several different mathematical ur-themes are often combined in a proof of a theorem. The interaction of ur-jokes is another type of "clothing".

(5) One important style of humor is the spotting of an unexpected or nonstandard type of "knob" on a familiar situation or entity, and the "twiddling" of that knob with unexpected consequences. Thus, consider the remark "If Harvard Square had been Harvard Cube, Briggs & Briggs would have been Briggs & Briggs & Briggs." You don't have to know that Briggs & Briggs is a music store in Harvard Square in order tc) get this joke, which involves the perception (and subsequent twiddling) of a "numbers knob", in which the first surprise is the unexpected slippage "square cube", which implies that in some sense, 2 has slipped to 3, and the s,_- cond surprise is the slippage "B & B B & B & B", which supposedly "follows from" the original one by some kind of "logic".

(6) Another important style of humor is the frame blend, exemplified by the two "Off the Leash" cartoons by W. B. Park, reproduced on the back of this sheet. In a frame blend, elements of two distinct situations or scenarios are combined to yield an (ostensibly) plausible new situation, but one that is actually totally implausible. A question that arises for frame Lends is: Why were these elements, and not others, selected for blending? What happens if one "twiddles knobs" on the given frame blend, adjusting the relative amounts of Frame I and Frame 2?

(7) There is a close kinship between frame blends and analogy. In a certain sense, a frame blend can be considered to be an imperfect or poor analogy, or conversely, an analogy can be considered to be a perfect frame blend.